\documentclass[useAMS, usenatbib]{mn2e}
\usepackage{epsf}
\usepackage{graphicx}

\newcommand{\msun}{\mathrm{M}_\odot}
\newcommand{\rsun}{\mathrm{R}_\odot}

\title[Braking rate of NN~Ser]{Detection of a period decrease in NN~Ser with ULTRACAM: evidence for strong magnetic braking or an unseen companion?} 
\author[C. S. Brinkworth et al.]{C. S. Brinkworth$^{1,2}$\thanks{E-mail: csb@ipac.caltech.edu
(CSB)}, T. R. Marsh$^{3}$, V. S. Dhillon$^4$, C. Knigge$^{2}$\\
$^{1}$Spitzer Science Center, California Institute of Technology M/C 220-6, 1200 E. California Blvd, Pasadena, CA 91125, USA \\ 
$^{2}$Dept of Physics \& Astronomy, University of Southampton, Highfield, Southampton, SO17 1BJ, UK\\ 
$^{3}$Department of Physics, University of Warwick, Coventry, CV4 7AL, UK \\
$^{4}$Department of Physics \& Astronomy, University of Sheffield, Sheffield, S3 7RH, UK }

\begin{document}

\date{Accepted ???? Received ????; in original form ????}

\pagerange{\pageref{firstpage}--\pageref{lastpage}} \pubyear{2002}

\maketitle

\label{firstpage}

\begin{abstract}
 We present results of high time resolution photometry of the
eclipsing pre-cataclysmic variable NN~Ser. NN~Ser is a
white dwarf/M dwarf binary with a very low-mass secondary star ($\sim
0.2\,\rm{M_{\odot}}$). We observed 13 primary eclipses of NN~Ser using the
high-speed CCD camera ULTRACAM and derived times of mid-eclipse, from
fitting of light curve models, with uncertainties as low as 0.06\,s. The
data show that the period of the binary is decreasing, with an average rate of $\dot{P} =  (9.06\pm0.06) \times10^{-12}\,\rm{s/s}$, which has increased to a rate of $\dot{P} = (2.85\pm0.15) \times10^{-11}\,\rm{s/s}$ over the last 2 years. These rates of period change appear difficult to reconcile with any models of orbital period
change. If the observed period change reflects an angular momentum loss, the average loss rate ($\dot{J} = 1.4 \pm0.6 \times 10^{35}$\,ergs) is consistent with the loss rates
(via magnetic stellar wind braking) used in standard models of close
binary evolution, which were derived from observations of much more
massive cool stars. Observations of low-mass stars such as NN~Ser's
secondary predict rates of $\sim$100 times lower than we
observe. The alternatives are either magnetic activity-driven changes
in the quadrupole moment of the secondary star \citep{ap92} or a light
travel time effect caused by the presence of a third body in a long
($\sim$\,decades) orbit around the binary. We show that Applegate's
mechanism fails by an order of magnitude on energetic grounds, but that the presence
of a third body with mass $ 0.0043\,\rm{M}_{\odot} < \rm{M}_{3} <
0.18\,\rm{M}_{\odot}$ and orbital period 30 $< \rm{P}_{3} <$ 285 years
could account for the observed changes in the timings of NN~Ser's
mid-eclipses. We conclude that we have either observed a genuine
angular momentum loss for NN~Ser, in which case our observations pose
serious difficulties for the theory of close binary evolution, or we
have detected a previously unseen low-mass companion to the binary.

\end{abstract}

\begin{keywords}
binaries: eclipsing -- stars: individual: NN Ser -- stars: evolution -- stars: fundamental parameters -- planetary systems
\end{keywords}

\section{Introduction}
The evolution of close binary systems is governed by angular momentum
(AM) loss, driven by a combination of gravitational radiation
(\citealt{kmg62}; \citealt{fa71}), which is dominant for periods
$P_{orb} <$ 3\,h,  and magnetic braking \citep{vz81}, which dominates
for $P_{orb} >$ 3\,h. Gravitational radiation is relatively well
understood, with the angular momentum loss rates simply governed by the masses and
separation of the components of the binary system. Magnetic braking,
however, is a more complicated issue. The mechanism is driven by the
magnetic field and stellar wind of one of the binary components. As
mass is driven off in the stellar wind, the ionised particles are
forced to co-rotate with the field lines out to the Alfv\'{e}n
radius. This draws angular momentum away from the star, effectively exerting a
braking force to slow its spin. In close binaries, the donor star is
tidally locked to the primary, so the angular momentum loss cannot act
to slow the spin period of the secondary alone. Instead, the angular
momentum is drawn from the binary orbit, causing it to shrink and the
orbital period of the binary to decrease. The rate of angular momentum loss by
magnetic braking is governed by the mass, radius and angular momentum
of the magnetically active star, but there is also evidence to suggest
that the AM loss rate saturates for low-mass stars above a certain
value of angular momentum \citep{spt00}. This has led to a major revision in
magnetic braking loss rates for binaries with low-mass secondaries,
such as cataclysmic variables.

Cataclysmic variable stars (CVs) are mass-transferring binary systems
comprising a white dwarf primary and a low-mass main-sequence
secondary. The majority have orbital periods between 1.3 and 10 hours, and
their evolution is governed, as with all close binaries, by angular momentum loss
from the system (see \citealt{w95} for a comprehensive review). CVs
are useful for testing close binary evolution, as any theory is constrained
by 2 major features in the distribution of CV periods: the period gap
and the value of the period minimum. The period gap is a dearth of
systems with periods between 2 and 3 hours. Standard CV theory
explains this gap by assuming that magnetic braking is cut off at
P\,$\sim$\,3\,h as the secondary becomes fully convective (there is no
longer a radiative/convective boundary to anchor the magnetic field,
so it either dissipates or is rearranged, resulting in a lowered stellar wind).  Mass transfer ceases until the system evolves to a period of $\sim$\,2\,h and gravitational radiation becomes strong enough to drive mass
transfer, re-populating the period distribution below
the gap.

The value of the minimum period is governed entirely by the angular momentum loss
rate for short-period systems. Under the standard theory, systems
below the period gap are driven by gravitational radiation alone, which implies that the
minimum period should be at $P_{min} = 1.1$\,h. In fact the observed
cut-off is at about $P_{min} \simeq 1.3$\,h, suggesting that gravitational radiation alone
is not strong enough to reproduce the observed value of $P_{min}$.

The standard model was developed by extrapolation from studies of braking rates of solar-type stars in clusters \citep{rvj83, sr83}. However, a recent dramatic increase in the amount of data available for stars in these clusters \citep[see][for reviews]{st97,
kr97, rm00} has shown that this extrapolation to lower masses appears
to be totally wrong - low-mass stars retain more of their angular momentum than
their higher-mass counterparts. This means that the new suggested $\dot{J}$ is anything between $10$ and $10^4$ times
smaller than assumed in the majority of CV studies. Importantly, there is also no evidence for
a cut-off in magnetic braking as the secondary becomes fully
convective \citep{aps03}, so the new data offer no explanation for the
existence of the period gap.

We therefore need a way to directly measure the angular momentum loss
rates of CVs in order to test the standard vs. reduced magnetic braking models. One way of
doing this is to measure mid-eclipse timings of eclipsing binary
systems to find the period change of the system and calculate the angular momentum
loss. The period changes are tiny - of order 5 $\times 10^{-4}$ seconds
per year - so this is difficult to do in CVs due to contamination of
the light curve by the accretion processes. In order to overcome this problem, we have used a non
mass-transferring pre-CV, NN~Ser. NN~Ser is ideally suited for this study as, in addition to the uncontaminated light curve, the system also displays deep primary eclipses that give rise to very sharp ingress and
egress features.  Given the high-time resolution of ULTRACAM, we
are able to estimate the times of mid-eclipse to an accuracy of $\sim
0.15$\,s.

NN~Ser is a white dwarf\,/\,M dwarf binary system with an extremely
low-mass (M $\sim 0.15\,\rm{M_{\odot}}$), and therefore fully convective
secondary star. The system was first studied in detail by \citet
{ha89}, who identified it as a deeply eclipsing ($ >4.8$\,mag)
pre-cataclysmic variable with a strong reflection effect of $\sim 0.6$\,mag, and an orbital period of 0.13 days. \citet{wm91} used
low-resolution $\it{IUE}$ spectra to derive the system parameters, which were
refined by the radial velocity study of \citet{ca94} to give the
values in Table~\ref{tab:syspar}. The most recent study by \citet{ha04} combines
high-speed photometry from the Multi-Channel Multi-Colour Photometer (MCCP) with VLT trailed
photometry and phase-resolved spectroscopy. This allows them to put
good constraints on the temperature of the secondary star. They also
attempt to derive accurate values for the radii and masses of the
system components, but they failed to detect
the secondary eclipse for NN~Ser, which caused them to derive a binary
inclination of $i = 84.6^{\circ} \pm 1.1^{\circ}$. We have detected
the secondary eclipse in our ULTRACAM data (see Fig.~\ref{fig:nnserlc}), and our
preliminary modelling indicates that the true inclination is $i \sim
88^{\circ}$. Full results of our modelling will be the subject of a
future paper, but we conclude from our initial results that
\citet{ha04} have overestimated the radius (and therefore the mass) of
the secondary star by $\sim 15\% $. Nevertheless, we carry out all of
our analysis for all values of mass and radius included in the
uncertainties given by \citet{ca94} and \citet{ha04}.  All four studies also give eclipse timings (listed in Table~\ref{tab:ecltimes}) which we have used to extend our baseline for measuring the period
change to $\sim$ 15 years. 

\begin{table}
\caption{System parameters of NN~Ser. RD = M dwarf secondary star} \label{tab:syspar}
\begin{tabular}{lll} \hline
& Catal\'{a}n et al. 1994 & Haefner et al. 2004 \\
& Wood \& Marsh, 1991 & \\ \hline
Binary sep. & 0.95 $\pm$ 0.025 R$_{\odot}$ & 0.9543 $\pm$ 0.0233R$_{\odot}$\\ 
Inclination & $84^{\circ} < i < 90^{\circ}$ & 84.6$^{\circ}\pm$ 1.1$^{\circ}$\\ 
Mass ratio & 0.18 $< q <$ 0.23 & 0.2778 $\pm$ 0.0297\\ 
WD mass & 0.57 $ \pm$ 0.04 $\rm{M}_{\odot}$ & 0.54 $\pm$ 0.05 $\rm{M}_{\odot}$\\ 
RD mass & 0.1 $< \rm{M}_{\odot} <$ 0.14 & 0.150 $\pm$ 0.008 $\rm{M}_{\odot}$\\ 
WD radius & 0.017 $< \rm{R}_{\odot} < 0.021$ & 0.0189 $\pm$ 0.0010 $\rm{R}_{\odot}$ \\ 
RD radius & 0.15 $< \rm{R}_{\odot} <$ 0.18 & 0.174 $\pm$ 0.009 $\rm{R}_{\odot}$\\
WD temp.  & 55000K $\pm$ 8000K & 57000K $\pm$ 3000K\\ 
RD temp. & 2900K $\pm$ 150K & 2950K $\pm$ 70K\\ 
RD irr. temp. & 5650K $< \rm{T} <$ 8150K & 7125K $\pm$ 200K\\
RD spec. type & M4.7 - M6.1 & M4.75 $\pm$ 0.25 \\
Distance & 356 pc $< d < $ 472 pc & 500 $\pm$ 35 pc \\ \hline
\end{tabular}
\end{table}

\section{Data acquisition}
\begin{figure}
\begin{center}
\includegraphics[width=3.4in]{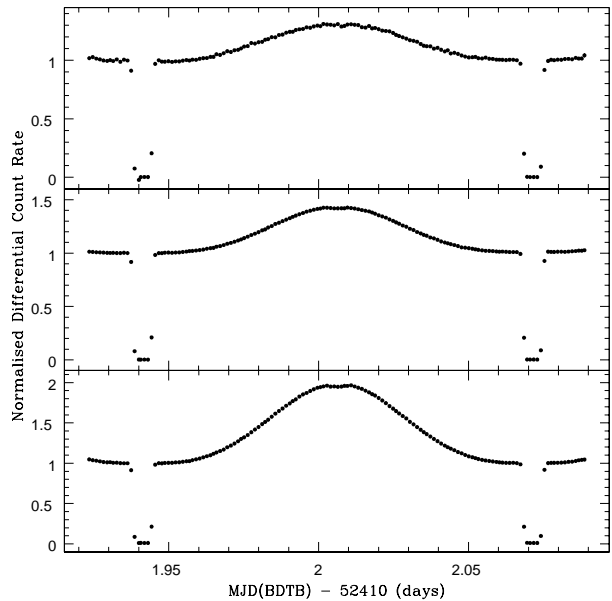}
\end{center}
\caption{Differential light curves for NN~Ser, taken simultaneously in
the u', g' and r' Sloan filters from top to bottom respectively. The light curves are binned by a factor of 43. The
hump in the light curve is caused by the reprocessing of light from
the WD by the cool secondary star. A shallow secondary eclipse can be detected at
the top of the reflection hump in the r' and g' bands.} \label{fig:nnserlc}
\end{figure}

The data were taken with the ultra-fast, triple-beam CCD camera,
ULTRACAM (see \citet{dm01} for a review). We used the camera in
conjunction with the 4.2\,m William Herschel telescope at the ING to
observe NN~Ser simultaneously in the Sloan u', g' and either r', i' or
z' bands. We set a time resolution of $\sim$\,2.06\,s to ensure a high signal-to-noise ratio in all wavebands. The observations were taken over a period of 2 years, in
May 2002 -- 2004, and during those runs we observed 13 primary eclipses
of the system. We were also able to observe a number of secondary
eclipses as the white dwarf transited the secondary star (see Fig.~\ref{fig:nnserlc}). The pixel size for the 3 ULTRACAM CCDs is 13\,$\mu$m, with
a scale of 0.3''/pix. Readout noise is 3.10 -- 3.40\,e, depending on the
CCD, while the gain is 1.13 -- 1.20\,e/ADU.Each ULTRACAM frame is time-stamped to a relative
   accuracy of better than 50~$\mu$s and an absolute accuracy of
   better than 1 millisecond using a dedicated GPS system.  A full list of observations and observing conditions is given in Table~\ref{tab:obslog}. The weather in May 2004 was particularly variable, leading to larger errors in our measured times for that epoch. 

\begin{table}
\caption{Observation log for ULTRACAM observations of NN~Ser} \label{tab:obslog}
\begin{tabular}{cccc} \hline
Date & Filters & No.      & Conditions \\
     &         & eclipses &    \\ \hline
17/05/2002 & u'g'r' & 2 & good, seeing $\sim$1.2'' \\
18/05/2002 & u'g'r' & 1 & variable, seeing 1.2'' - 2.4''\\
19/05/2002 & u'g'r' & 1 & fair, seeing $\sim$2'' \\
20/05/2002 & u'g'r' & 1 & fair, seeing $\sim$2'' \\
19/05/2003 & u'g'z' & 1 & variable, seeing 1.5'' - 3'' \\
21/05/2003 & u'g'i' & 1 & excellent, seeing $\sim$1'' \\ 
22/05/2003 & u'g'i' & 1 & excellent, seeing $<$1''\\
24/05/2003 & u'g'i' & 1 & good, seeing $\sim$ 1.2''\\
03/05/2004 & u'g'i' & 3 & variable, seeing 1.2'' - 3.2'' \\
04/05/2004 & u'g'i' & 1 & variable, seeing 1.2'' - 3'' \\  \hline
\end{tabular}
\end{table}

The data were reduced using the ULTRACAM pipeline software written by
TRM. Differential photometry was performed on the target, with respect to a
nearby, bright, non-variable comparison star.

\begin{figure}
\begin{center}
\includegraphics[width=2.3in, angle=270]{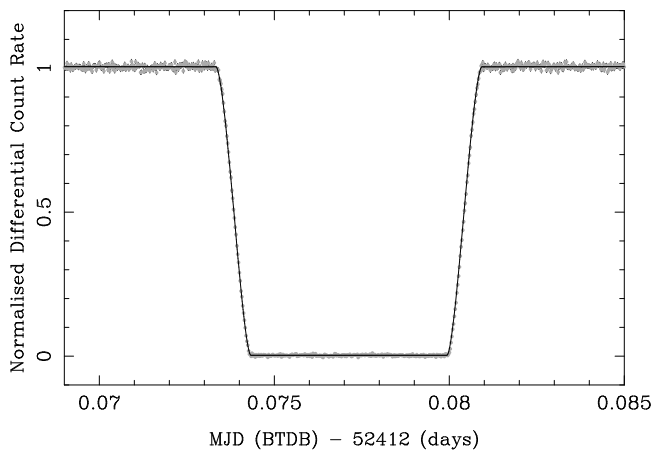}
\end{center}
\caption{Differential light curve for NN~Ser in g' with light curve model overplotted.} \label{fig:lcmodel}
\end{figure}


\begin{table*}
\caption{Measured times of mid-eclipse for each of the 13 observed primary eclipses of NN~Ser. Cycle numbers are counted from first measured eclipse in the literature \citep{ha89}. Times were measured for all 3 wavebands simultaneously observed with ULTRACAM and are given in MJD(BTDB), i.e. MJD shifted to the solar system barycentre and corrected for light travel time. The red filter varied between nights, so the filter used is listed in the final column. The poor observing conditions during eclipse cycle 44474 led to the loss of data in the u' band. } \label{tab:ecltimes}
\begin{tabular}{cccccccc} \hline
Cycle  & u' eclipse     & uncertainty     & g' eclipse     & uncertainty     & r'/i'/z' eclipse & uncertainty     & red    \\ 
Number & time           &   1 $\sigma$        & time           &  1 $\sigma$         & time             &   1 $\sigma$        & filter \\ \hline
38960      & 52411.9470588 & 0.0000020 & 52411.9470564 & 0.0000005 & 52411.9470577   & 0.0000010 & r'     \\
38961      & 52411.0771385 & 0.0000016 & 52412.0771385 & 0.0000005 & 52412.0771383   & 0.0000010 & r'     \\
38968      & 52412.9876761 & 0.0000030 & 52412.9876977 & 0.0000008 & 52412.9876721   & 0.0000013 & r'     \\
38976     & 52414.0283427 & 0.0000030 & 52414.0283394 & 0.0000006 & 52414.0283379   & 0.0000016 & r'     \\  
38984     & 52415.0689716 & 0.0000025 & 52415.0689810 & 0.0000007 & 52415.0689795   & 0.0000016 & r'     \\
41782   & 52779.0331646 & 0.0000021 & 52779.0331696 & 0.0000010 & 52779.0331362   & 0.0000100 & z'     \\ 
41798   & 52781.1144524 & 0.0000015 & 52781.1144513 & 0.0000006 & 52781.1144567   & 0.0000014 & i'     \\ 
41806   & 52782.1550904 & 0.0000021 & 52782.1550929 & 0.0000006 & 52782.1550948   & 0.0000011 & i'     \\ 
41820   & 52783.9762155 & 0.0000022 & 52783.9762151 & 0.0000007 & 52783.9762110   & 0.0000020 & i'     \\
44472   & 53128.9486787 & 0.0000070 & 53128.9486778 & 0.0000040 & 53128.9486611   & 0.0000800 & i'     \\ 
44473   & 53129.0787555 & 0.0000027 & 53129.0787597 & 0.0000022 & 53129.0787487   & 0.0000050 & i'     \\
44474   & no data        & n/a      & 53129.2088356 & 0.0000020 & 53129.2088355   & 0.0000027 & i'     \\
44480   & 53129.9893197 & 0.0000050 & 53129.9893229 & 0.0000025 & 53129.9893148   & 0.0000040 & i'     \\ \hline

\end{tabular}
\end{table*}

\section{Analysis \& Results}

All MJD times were corrected to Barycentric Dynamical Time (TDB), then additionally
corrected for light travel time to the solar system
barycentre. All times are therefore listed in MJD(BTDB).  
In order to measure accurate eclipse times, we needed a model of the 
eclipse of the white dwarf, which we calculated as follows. We defined the
two stars by their radii relative to the separation of the binary. Since 
we allowed for tidal deformation of the M dwarf (but not the white dwarf),
the radius of the M dwarf was measured from its 
centre of mass towards the white dwarf. Apart from the relative radii, we 
also require the binary mass ratio and inclination, stellar effective 
temperatures and linear limb darkening coefficients to define our model 
binary. The two stars were divided into many small elements. The 
temperatures of the elements covering the M dwarf were set, accounting 
for incident flux from the white dwarf by adding fluxes so that
\[ \sigma {T'}_{2}^{4} = \sigma T_{2}^{4} + F_{\mathrm{irr}} , \]
where $\sigma$ is the Stefan-Boltzmann constant and $F_\mathrm{irr}$ is 
the flux incident on the secondary accounting for the projection effects and the distance from 
the white dwarf. The surface brightness of each element was then set 
assuming black-body spectra, and given the effective wavelength of the 
filter in question. Once the surface brightnesses were set, the model 
light-curves were computed by summing over all elements, testing for 
which were in view and not eclipsed and accounting for their projected areas.
The eclipse by the M dwarf was computed, allowing once again for tidal 
distortion. Our assumption of black-body spectra for the two stars is physically 
unrealistic, but for the eclipse times of this paper, the key element 
is to have a model that can match the shape of the primary eclipse, which ours does well (Fig.~\ref{fig:lcmodel}).
The timings and associated errors for the mid-eclipses in all of the wavebands are given in Table~\ref{tab:ecltimes}. The errors on our mid-eclipse timings are 
typically $\sim$\,0.15\,s but as low as 0.06\,s when conditions are good. 

\begin{table}
\caption{Previous eclipse times of NN~Ser (1) Haefner (1989); (2) Wood \& Marsh (1991); (3) Pigulski \& Michalska (2002); (4) Haefner et al. (2004).} \label{tab:oldecltimes}
\begin{tabular}{lll} \hline
Time of mid-eclipse & Cycle & Reference \\
MJD(BTDB) & Number & \\ \hline
47344.025(5) & 0 & 1 \\
47703.045744(2) & 2760 & 4 \\
47703.175833(6) & 2761 & 4 \\
47704.216460(3) & 2769 & 4 \\
47705.127023(3) & 2776 & 4 \\
47705.257115(7) & 2777 & 4 \\
47712.28158(15) & 2831 & 2 \\
47713.32223(15) & 2839 & 2 \\
48301.41420(15) & 7360 & 2 \\
51006.0405(2)   & 28152 & 4 \\
51340.2159(2)   & 30721 & 4 \\
51666.9779(4)   & 33233 & 3 \\ \hline
\end{tabular} 
\end{table}

\begin{figure*}
\begin{center}
\includegraphics[height=5.5in, angle=0]{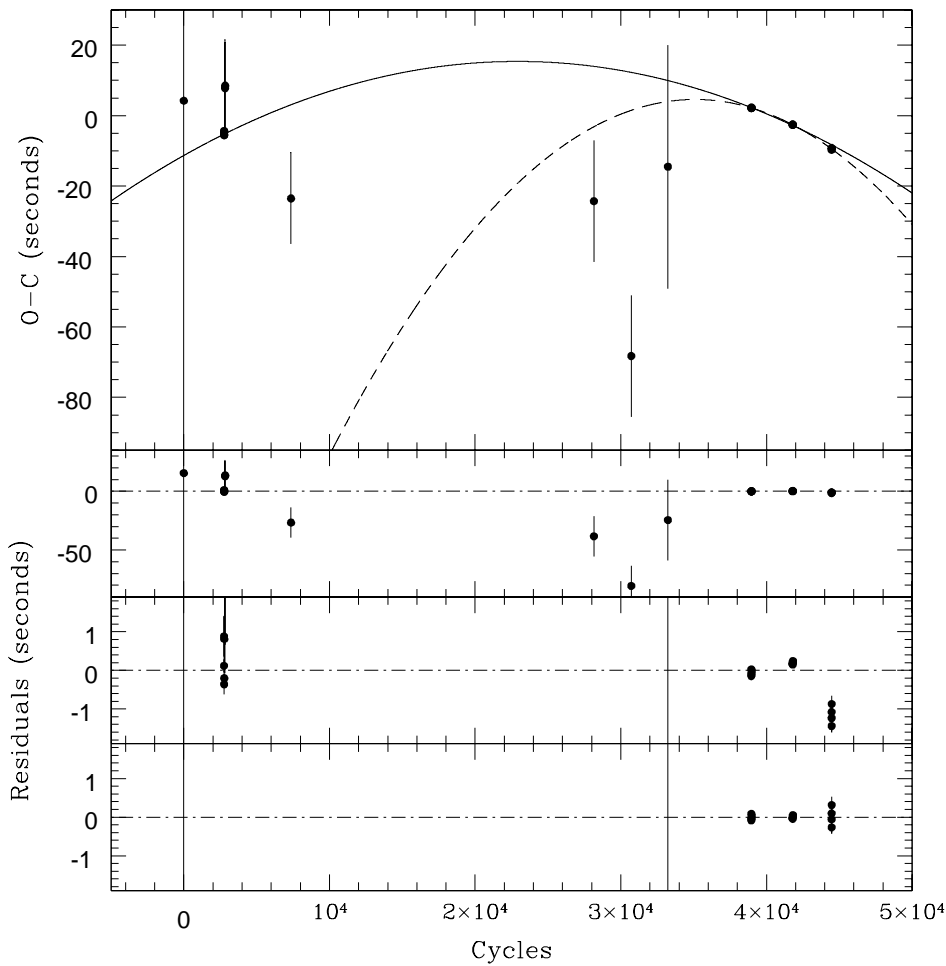}
\end{center}
\caption{The upper plot is an O-C diagram showing the period change in NN~Ser. A linear fit has been subtracted from the data. The solid line is a fit through all the data (average rate of period change), while the dashed line is a fit through the ULTRACAM data only (current rate of period change). The lower three plots are (from uppermost): residuals after a fit through all the data is subtracted, showing all the points; residuals after the fit through all the data is subtracted, zoomed in on the ULTRACAM points; residuals after the fit through the ULTRACAM data is subtracted, zoomed in on those points.} \label{fig:o-c}
\end{figure*}

The times of mid-eclipse in the g' band were plotted against cycle number. We
found that all 13 of the ULTRACAM points except for one (cycle 38968) fell on a straight line, and that the one discrepant point showed a
time shift of 2.06\,s - exactly the same timing as one exposure. We noted from the logs that we had GPS problems during this exposure run, therefore concluded that the GPS timestamp had slipped by one exposure
for that point, and corrected it by 2.06\,s to bring it in line with the
other points. Old eclipse timings from the literature (\citealt{ha89};
\citealt{wm91}; \citealt{pm02}; \citealt{ha04}; Table~\ref{tab:oldecltimes}) were then added
to the plot. The residuals after subtracting a straight-line fit can
be seen in Fig.\ref{fig:o-c}. We derive a best-fit linear ephemeris from all the available data for NN~Ser as 

\[\rm{MJD(BTDB)} = 47344.0246049(14) + 0.130080144430(36)E,\]

where the quoted uncertainties are the 1 sigma uncertainties in the fit. We derive a best-fit quadratic ephemeris as:

\[\rm{MJD(BTDB)} = 47344.0244738(16) + 0.13008017141(17)E \]
\[                                - 5.891(36) \times 10^{-13} \rm{E}^{2}.\]

The data, with eclipse times t$_{E}$, were fit with a parabola of form:

\begin{equation}
t_{E} = T_{0} + AE + BE^{2}.
\end{equation}

\noindent
The rate of period decrease can then then be found using

\begin{equation}
\dot{P} = \frac{2B}{P}.
\end{equation}

We found that the rate of period change over the 15 years of observations is increasing, so we fit all of the data to find an average rate of period change, and just the ULTRACAM data to find the current rate. 

\noindent
The angular momentum of the system as a whole is given by

\begin{equation}
J = \left(\frac{Ga}{M}\right)^{1/2} M_{1} M_{2},
\end{equation} 

\noindent
 where $M_{1}$, $M_{2}$ and M are the primary, secondary and total
 masses respectively. Combining this with Kepler's third law,

\begin{equation}
\frac{4\pi^{2}}{P^{2}} = \frac{GM}{a^{3}},
\end{equation}

\noindent
we find that, for a detached system (where $M_{1}$, $M_{2}$ and M are constant), 

\begin{equation}
\frac{\dot{J}}{J} = \frac{2}{3}\frac{B}{P^{2}}.
\end{equation}

\noindent
For NN~Ser, our measured value for the average rate of period change is

\[\dot{P}_{av} = 9.06 \times 10^{-12} \pm 0.06 \times 10^{-12}\,\rm{s/s}\]

\noindent 
and for the current rate of period change

\[\dot{P}_{cur} = 2.85 \times 10^{-11} \pm 0.15 \times 10^{-11}\,\rm{s/s}.\]

Taking $0.1\,\rm{M_{\odot}} \leq \rm{M_{2}} \leq 0.14\, \rm{M_{\odot}}$ and $0.15 \,\rm{R_{\odot}} \leq \rm{R_{2}} \leq 0.18\,\rm{R_{\odot}}$ \citep{ca94}, these correspond to angular momentum loss rates of 

\[0.84 \times 10^{35} \leq \dot{J}_{av} \leq 2.09 \times 10^{35}\,\rm{ergs}.\]

and

\[2.52 \times 10^{35} \leq \dot{J}_{cur} \leq 6.87 \times 10^{35}\,\rm{ergs}.\]

\noindent
where the relatively large allowed range is caused by the uncertainties in the system parameters, where we have assumed that the system
parameters are independent of each other. Obviously this will
overestimate the size of the uncertainty in our measured angular momentum loss rate
for any one value of secondary mass. When carrying out the analysis of
period loss mechanisms in Section 4, we have used a more realistic
approach, relating the secondary radius to its mass, using the M-R
relation for secondaries in binary stars given in \citet{gs98} and
calculating the resultant separation of the binary. This then gives a
range of values for the angular momentum change that are specific to 
each value of secondary mass.

\section{Discussion - Mechanisms for period change}
Period changes in binary systems are generally due to one of three mechanisms:

\begin{enumerate}
\item Applegate's (1992) mechanism, where period changes are caused by coupling between the binary period and changes in the shape of the secondary star; 
\item the presence of a third body in a long orbit around the binary. This affects the light travel-time, which can be mis-interpreted as a change in the binary period. For example, as the binary moves towards the observer, the eclipses are seen to occur more frequently than when the binary is moving away.
\item a genuine angular momentum loss from the system;
\end{enumerate}

We show below that the most common cause of measurable change in binary periods -
Applegate's mechanism - cannot work for NN~Ser: the luminosity of
the secondary star is too low to provide the necessary energy. We also discuss the other two mechanisms in detail, along
with the ramifications for binary evolution.

\subsection{Applegate's mechanism}
\citet{ap92} proposed that orbital period modulations observed in many binary
stars could be induced by the gravitational coupling of the
binary orbit to variations in the shape of the magnetically active
secondary star. The shape changes are reflected in a change of quadrupole moment which 
leads to the change in period; no loss of angular momentum from the system is
necessary. The shape changes are presumed to be driven by solar-like magnetic
cycles. To avoid an excessive energy budget \citep{mp90}, 
\citet{ap92} proposed that the shape changes were driven by the re-distribution
of angular momentum within the secondary star. He showed that the energy
required was well within the capabilities of 4 out of 5 stars that he considered, and
was not far off the mark for the fifth, RS~CVn.

One of the reasons we chose NN~Ser was that it has a particularly low mass --
and therefore low luminosity -- secondary star, which should be incapable of
driving large period changes under \citet{ap92}'s model. We now consider this in
detail.

The observational fact we have to explain is the total period change, which in
NN~Ser amounts to $\Delta P = (- 4.26 \pm 0.03) \times 10^{-3}$\,s over the 15 years
from the MCCP to the ULTRACAM epoch. A period change
$\Delta P$ corresponds to a change in quadrupole moment $\Delta Q$, where
\begin{equation}
\frac{\Delta P}{P} = - 9 \left( \frac{R}{a} \right)^2 \frac{\Delta Q}{M R^2} \label{eq:deltap},
\end{equation}
where $M$ and $R$ are the mass and radius of the secondary star and $a$ is the
orbital separation \citep{ap87}. \citet{ap92} calculated the change in quadrupole moment by
considering the transfer of angular momentum from the inside of the star into a 
thin outer shell. This increases the oblateness of the shell and therefore its
quadrupole moment, at the expense of some energy. \citeauthor{ap92} used $M_s \sim
0.1\,\msun$ for the mass of the shell. This is immediately a problem in the case
of NN~Ser because the mass of the secondary star is only $0.15 \,\msun$, and so
$0.1\,\msun$ is not in any sense a ``shell''. We therefore generalised
\citeauthor{ap92}'s work as follows. We split the star into an inner ``core'',
denoted by subscript 1 and an outer shell denoted by subscript 2. Angular
momentum is transferred from the core to the shell leading to changes in their
angular frequencies of $\Delta \Omega_1$ and $\Delta \Omega_2$ such that
\begin{equation}
I_1 \Delta \Omega_1 + I_2 \Delta \Omega_2 = 0 \label{eq:deltaj},
\end{equation}
where $I_1$ and $I_2$ are the moments of inertia and given by integrals over 
radius of a series of thin shells of the form
\begin{equation}
I = \frac{2}{3} \int R^2 \,\mathrm{d} M .
\end{equation}
The change in angular frequencies changes the oblateness and therefore
quadrupole moments of the core and shell leading to an overall change in
quadrupole moment of 
\begin{equation}
\Delta Q =  
Q'_1 \left[2 \Omega_1 \Delta \Omega_1 + \left(\Delta \Omega_1\right)^2\right] +
Q'_2 \left[2 \Omega_2 \Delta \Omega_2 + \left(\Delta \Omega_2\right)^2\right]
\label{eq:deltaq},
\end{equation}
where the $Q'$ coefficients are given by integrals over shells of the form
\begin{equation}
Q' = \frac{1}{9} \int \frac{R^5 \,\mathrm{d} M}{G M(R)}, 
\end{equation}
where $M(R)$ is the mass inside radius $R$. These equations follow from 
Eq.~25 of \citet{ap92}. 

For a given period change, mass and radius Eq.~\ref{eq:deltap} gives the
change in quadrupole moment. We then use Eqs~\ref{eq:deltaj} and \ref{eq:deltaq}
to solve for $\Delta \Omega_2$, and therefore for the angular momentum
transferred, $\Delta J = I_2 \Delta \Omega_2$. This then leads to the energy
change from \citet{ap92}'s Eq.~28:
\begin{equation}
\Delta E = \Omega_{dr} \Delta J + \frac{1}{2} \left(\frac{1}{I_1} +\frac{1}{I_2}\right)
  \left(\Delta J\right)^2,
\end{equation}
where $\Omega_{dr} = \Omega_2 - \Omega_1$ is the initial differential rotation.

In order to solve the equations above, one needs first to know the run of
density with radius. We calculated this from the Lane-Emden equation for an $n =
1.5$ polytrope as an approximation to the fully convective secondary star.

\begin{figure}
\begin{center}
\includegraphics[angle=270,width=0.9\columnwidth]{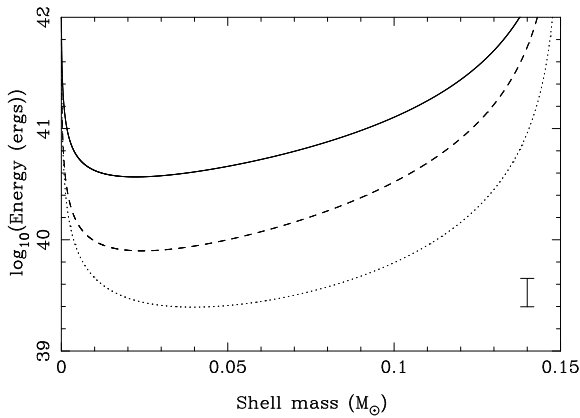}
\end{center}
\caption{A plot of the energy required to effect the period change in NN~Ser
  using \protect\cite{ap92}'s mechanism as a function of assumed shell mass.
The dotted line shows \protect\citeauthor{ap92}'s original calculation.
The solid line shows our calculation, integrating over shells and allowing
for the quadrupole moment of the inner core of the star. The dashed line shows
the result if we ignore the quadrupole moment of the core. The error bar in the
lower-right shows the energy available to effect the change in NN~Ser.}
\label{fig:applegate}
\end{figure}

\begin{figure}
\begin{center}
\includegraphics[width=3.3in]{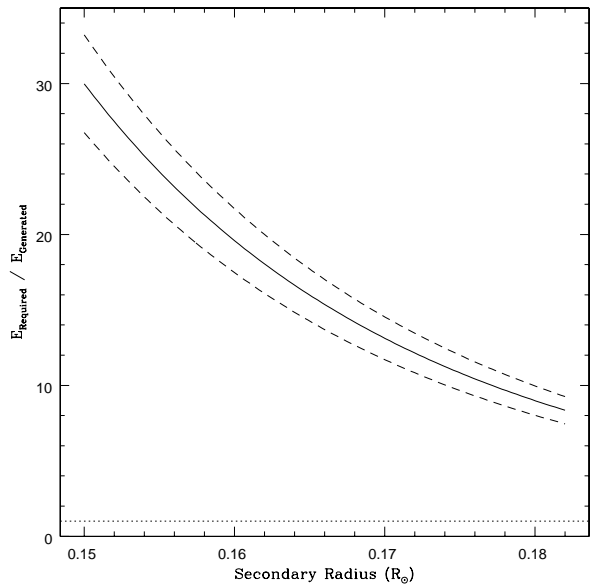}
\end{center}
\caption{Ratio of energy required for Applegate's mechanism over energy generated by the secondary star, vs secondary radius. The dashed lines show the uncertainty in energy ratio due to the uncertainties in the observed period change and the temperature of the secondary. The dotted line is at a ratio of 1, i.e. Applegate's mechanism for period change is only possible below this line.} \label{fig:applegate_sec}
\end{figure}

In Fig.~\ref{fig:applegate} we show the results of these calculations as a
function of the shell mass for $M = 0.15\,\msun$, $R = 0.174 \, \rsun$
and $\Delta P = - 0.00426 \,$s. 
\citeauthor{ap92} used a value of $\Omega_{dr} = \Delta \Omega_2$, on the basis
that one would expect the initial differential rotation to be of a similar order
of magnitude as the changes. In order to arrive at a minimum energy, we have
assumed that $\Omega_{dr} = 0$; had we used instead $\Omega_{dr} = \Delta
\Omega_2$, the energies would be increased by a factor $\sim 2$. Even without
this factor, the figure shows that we need at least $\sim 4 \times
10^{40}\,$ergs  to drive the observed period change. The luminosity of the secondary
star is $L_{2} = 4 \pi R^{2} \sigma T_{eff}^{4}$, which, for NN~Ser's
secondary with $2880\,\rm{K} < T_eff < 3020\,\rm{K}$, gives the energy available over the 
15 years of observations as
\[2.5 \times 10^{39} \, \mathrm{ergs} \leq E_{2} \leq  4.5 \times 10^{39}\, \mathrm{ergs}.\] 
This range, which is indicated in the lower-right of Fig.~\ref{fig:applegate}
fails by a factor of about 10 to match the energy we calculate, although it does
just tally with a calculation based upon \citeauthor{ap92}'s original
calculations (dotted line). A plot of the ratio of our value of minimum energy required to drive Applegate's mechanism over the energy available in 15 years vs the radius of the secondary star is shown in Fig~\ref{fig:applegate_sec}. It can be seen that the ratio of energy required over energy available is well above 1 for all values of secondary star radius (and therefore mass), i.e. for the system parameters derived by \citet{ca94} and \citet{ha04}, NN~Ser's secondary star is not capable of generating enough energy to drive Applegate's mechanism.

Our increased estimate is a result of the differences in our approach compared
to \citet{ap92}'s. First we generalise his thin shell approximation by
integration over finite ranges of radii. Second, in Eq.~\ref{eq:deltaq}, there
is a positive contribution from the shell, but also a negative contribution from
the core which \citeauthor{ap92} did not include and which balances the shell to
a considerable extent. The dashed line in Fig.~\ref{fig:applegate} shows the
effect of ignoring the core's contribution to the energy in our
calculation. This line converges towards \citeauthor{ap92}'s (dotted line) at
small shell masses. At large shell masses the effect of the variation in density with
radius and finite shell thickness, which \citeauthor{ap92} did not include, are
important and explain the remaining difference. Ignoring the core cuts the
energy requirement by about a factor of 4, suggesting that, given the squared
dependence of energy on $\Delta J$ and hence $\Delta Q$, the core balances about
50\% of the quadrupole increase from the shell. \citet{ap92} already
recognised that his approximations must break down when the shell becomes a
significant fraction of the star's mass; the counter-balancing effect of the
core's quadrupole moment has not been pointed out before as far as we are aware.

While we have shown that the intrinsic luminosity of NN Ser's
secondary star is too low to drive \citet{ap92}'s mechanism, we note
that the secondary star is heavily irradiated by the hot, white dwarf
primary, as shown by the much higher temperature on the side facing
the white dwarf. In order to estimate how much of an effect this will
have on the secondary star, we compared the flux from the primary that
is intercepted by the secondary star with the secondary star's
intrinsic luminosity. We find that the intercepted flux from the white
dwarf is $\sim$\,13 times the intrinsic luminosity of the secondary, and
therefore, if more than 70\% of this were to be absorbed by the
secondary star, the extra energy provided could be enough to drive
\citet{ap92}'s mechanism. However, there are a number of reasons that
this should not be the case. Firstly, very little of the energy
absorbed at the stellar surface flows inward, since the opacity in
this region becomes very high, the stellar surface quickly heats up,
and so the heating luminosity is re-radiated
\citep{hr91,hr95}. Instead, the main effect of the irradiating flux is
to block the outflow of the radiation produced in the secondary star's
interior. The star will undergo a small expansion as some of the
blocked energy is stored as internal or gravitational energy, but
unlike a star undergoing isotropic irradiation, an anisotropically
irradiated star diverts the energy flow in the upper layers of the
convection zone to the unirradiated parts of its surface, efficiently
cooling the secondary star \citep{vn84,rz00}. In this case the energy
flow is decoupled from the mechanical and thermal structure, which can
still be considered as spherically symmetrical, hence the structure of
the secondary star below the convection zone is virtually unaffected
by the irradiating flux. The irradiating flux does not penetrate
deeply enough into the atmosphere to affect the deep boundary layers
that must be deformed to drive \citet{ap92}'s mechanism. Finally, if
the secondary star were absorbing $\sim$\,9 times its intrinsic energy from
the primary star, we should see some variability in the light curves
over the 3 years of observations with ULTRACAM. We see no evidence for
this as the light curves are consistent to within 3.5\% over that time.
 

We conclude that in the case of NN~Ser, \citet{ap92}'s quadrupolar distortion
mechanism falls short of being able to match the observed period change, although we note that there may yet be ways to affect the quadrupole distortion at less energy cost \citep{lr98}. We now look at alternative mechanisms of producing NN~Ser's period change. 

\subsection{Third body}
Apparent changes in the orbital periods of binary stars have often been
attributed to the light-travel time variation caused by third bodies, although
further observation usually reveals that this cannot be the case. However with
the relatively limited coverage to date, this is at least a possibility for
NN~Ser which we investigate in this section.

Changes in eclipse timings of binary stars do not necessarily indicate a genuine
change in the binary period. A third body in a long orbit around the binary can
cause small but significant changes in the light-travel time from the binary
system, which manifest themselves as strictly sinusoidal changes in the timings
of mid-eclipse. We are able to put constraints on the mass and period of any
third body which could cause the observed period change in NN~Ser by fitting all
possible sine waves to a plot of mid-eclipse timings vs cycle number. A function
of form
\[T = T_{0} + P_{orb}E + A_{3} \sin \left( \frac{2 \pi (E - E_{3})}{P_{3}}\right) \] 
was fitted to the plot for values of $P_{3}$ between 2 and 500000 days, where
$P_{orb}$ was kept fixed at the orbital period of the binary, $P_{3}$ is the
modulation period of the period change, $A_{3}$ is the amplitude of the period
modulation, and $E-E_{3}$ is a measure of the phase of the zero point of the
modulation with respect to the zero point of the binary period. As we are interested in the
minimum possible mass, we assumed that the inclination of the orbital plane of
the third body is aligned with the line of sight, i.e. sin $i =$ 1. This gave us
the values of A$_{3}$ for all possible modulation periods between 2 and 500000
days. From this, we were able to use Kepler's law and the observed luminosity of
NN~Ser to find the range of allowable masses of the third body which could cause
the observed period change in NN~Ser. The minimum possible mass comes from the
fact that we have not seen a reversal in the period change of NN~Ser.  The
minimum value of $P_{3}$ is therefore $\sim$ 30 years, which corresponds to a
minimum mass of $M_{3} =$ 0.0043\,M$_{\odot}$.

The maximum value of $M_{3}$ comes from the luminosity of the binary system in
eclipse. The luminosity of the third body must be equal to or less than the
observed mid-eclipse luminosity. This means that it must have a mass equal to or
less than that derived for the secondary star. If the maximum mass is
0.18M$_{\odot}$ then the maximum orbital period for any third body is $P_{3} =
1.04 \times 10^{5}$ days $\sim$ 285 years.

We therefore find that, on the basis of the current data at least, a low-mass
companion to the binary system could cause the observed changes in mid-eclipse
timings that we observe in NN~Ser, and that the long periods suggested by our
data would be able to accommodate NN~Ser's primary even before its evolution to
a white dwarf. Our results also indicate that measuring eclipse timings of
binary systems is potentially a very sensitive method of detecting extra-solar
planets in long-period orbits. We suspect however, that as in other instances,
further observations will rule out a third body.

\subsection{Comparison with angular momentum loss models}\
The period decrease we have measured in NN~Ser may also be explained by angular momentum loss from the binary system.
Angular momentum loss in CVs and pre-CVs is governed by two
mechanisms - gravitational radiation and magnetic braking. The rates
of AM loss caused by both mechanisms must be added together to find
the total AM loss for the system. We compare the inferred angular
momentum loss rate for NN~Ser (corresponding to the rate of period decrease) to both the values predicted by the standard
CV magnetic braking rate \citep{rvj83}, based on extrapolation from
studies of braking rates of solar-type stars in clusters, and to the reduced
magnetic braking rate \citep{spt00}, based on more recent data, for
which the angular momentum saturates at lower masses.  Under the standard model, the
angular momentum, J, decreases as $\dot{J} \propto  - \omega ^{3}$
\citep{sk72}, where $\omega$ is the angular velocity of the
star. However, the reduced braking model suggests that the angular
momentum loss is best modelled as $\dot{J} \propto - \omega ^{3}$ for
$\omega < \omega_{crit}$ and $\dot{J} \propto - \omega$ for $\omega >
\omega_{crit}$, where the threshold rate, $\omega_{crit}$ is much
lower than the rotation rates of CVs.  This means that the new
suggested $\dot{J}$ is anything between $10$ and $10^4$ times smaller
than assumed in the majority of CV studies.

If this is
correct, we require a large-scale revision of CV evolution, possibly
with systems staying at an approximately fixed period throughout their
lifetime rather than migrating from long to short periods. However,
such a model has significant problems when compared to observations,
particularly as the mass transfer rate should be much lower than seen
in the high accretion rate group of CVs known as novalike variables.

Both models were applied to CV studies by \citet{aps03}, hereafter APS03.

\subsubsection{Gravitational radiation}

We use the same expression for AM loss due to gravitational radiation
as used in APS03, although this was misquoted in their paper. The
correct expression is given by:
\begin{equation}
\left(\frac{dJ}{dt}\right)_{grav} = -\frac{32}{5}\frac{G^{7/2}}{c^{5}} a^{-7/2}M_{1}^{2}M_{2}^{2}M^{1/2}
\end{equation}
where M$_1$, M$_2$ and M are the white dwarf mass, secondary mass and total mass
respectively, and a is the binary separation given by Newton's form of Kepler's
third law $a = (GM/\omega^{2})^{1/3}$. For NN~Ser this gives a range of values
of 5.75 $\times 10^{32}\,\rm{ergs} < \dot{J}_{grav} <$ 1.74 $\times 10^{33}$\,
ergs, over 100 times smaller than required to drive our measured value of
$\dot{P}$ for NN~Ser.

\subsubsection{Standard magnetic braking model}

The standard model for magnetic braking in CVs is based upon studies
of the solar wind and and the rotation periods of solar-type stars in
open clusters (\citealt{wd67}; \citealt{sk72}; \citealt{ms87}). \citet{rvj83}
developed an empirical prescription that is still commonly used in CV
studies. This relationship is given by
\begin{equation}
\left(\frac{dJ}{dt}\right)_{mb} \approx -3.8 \times 10^{-30} M_{\odot} R_{\odot}^{4} m_{2} r_{2}^{\gamma} \omega^{3}\, \rm{ergs},
\end{equation}
where 0 $\leq \gamma \leq$ 4 is a dimensionless parameter and $\omega$
is the angular frequency of rotation of secondary star (= binary
period for CVs) in rad s$^{-1}$ . We applied this to 
NN~Ser to find the predicted
standard angular momentum loss rate for this pre-CV. The results can be seen in Fig.~\ref{fig:braking}. APS03 cut off the standard magnetic braking model at a secondary
mass of 0.3\,M$_\odot$ to satisfy the standard CV theory. This states
that as the secondary becomes convective, the magnetic field is no
longer locked to the stellar core and so either dissipates or is rearranged, cutting off the
magnetic braking mechanism. APS03 suggested that there is no evidence
for this cut-off, so we have not applied it here. We find that by ignoring the magnetic braking cut-off, this model can explain the observed loss rates seen in NN~Ser.


\subsubsection{Reduced magnetic braking model}

The more recently proposed model for angular momentum loss due to magnetic braking
was applied to CVs in APS03. Studies of the rotational periods of
low-mass stars (\citealt{qu98}; \citealt{cj94}; \citealt{kmc95};
\citealt{kr97}; \citealt{spt00}) all showed that the standard model
overestimates angular momentum loss rates for periods below 2.5 -- 5 days and that a
modification of the standard model was required for those high
rotation rates. APS03 modelled the modified angular momentum loss rates using a
prescription with the same functional form as that of \citet{spt00},
given by
\begin{equation}
\left(\frac{dJ}{dt}\right)_{mb} = -K_{w} \sqrt{\frac{r_{2}}{m_{2}}} \left\{ \begin{array}{ll}                                                                          \omega^{3} & \mbox{for } \omega \leq \omega_{crit} \\                                                              \omega\omega_{crit}^{2} & \mbox{for } \omega > \omega_{crit}                                                       \end{array}                                                                                                       \right.
\end{equation}
where $\omega_{crit}$ is the critical angular frequency at which the
angular momentum loss rate enters the saturated regime. The constant
K$_{w}$ = 2.7 $\times 10^{47}$ g cm s$^{-1}$ is calibrated to give the
known solar rotation rate at the age of the Sun \citep{ka88}. The
values of $\omega_{crit}$ were calculated from the values of
$\omega_{crit\odot}$ given in \citet{spt00} using the relationship
between $\omega_{crit}$ and convective turnover time, $\tau$ given by
\begin{equation}
\omega_{crit} = \omega_{crit\odot} \frac{\tau_{\odot}}{\tau}.
\end{equation}
The values of $\tau$ were taken from \citet{kd96}, assuming an age of 0.2\,Gyr. 

\begin{figure}

\begin{center}
\epsfxsize=3.4in \epsfbox{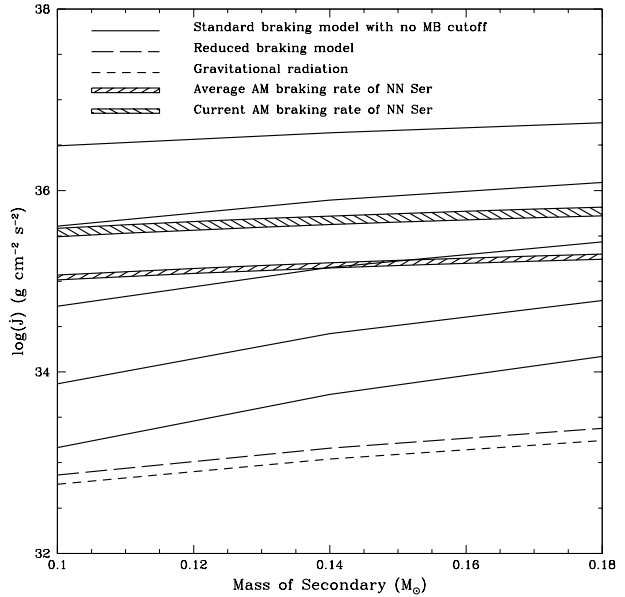}
\end{center}

\caption{Plot of the braking rates predicted by gravitational
radiation and the standard and reduced braking models for NN~Ser. The
different plots for the standard model are for different values of
$\gamma$ = 0,1,2,3,4, from the top down. The shaded region shows our
measured value of the braking rate of NN~Ser.}
\label{fig:braking}
\end{figure}

Again, the prescription was applied to NN~Ser. Results are shown in
Fig.~\ref{fig:braking}. Our plot differs significantly from the original plot in
APS03 due to their mis-calculation of the angular momentum loss due to
gravitational radiation. By applying the correct gravitational radiation loss
rate, the total angular momentum loss rates predicted by APS03 are within $\sim$
1 order of magnitude of the standard magnetic braking model rather than $\sim$ 2
orders of magnitude lower, as they originally claimed. However, this is still
too low to explain the loss rates seen in NN~Ser.

\section{Discussion}
We have found that only two mechanisms can explain the observed period change in
NN~Ser - either a genuine angular momentum loss from the system or an unseen
third body in orbit around the binary.  In the case of an angular momentum loss,
our observations show that the system is losing angular momentum at the rate
predicted by \citet{rvj83}, but only if we assume that magnetic braking is not
cut off as the secondary mass reaches 0.3\,$\rm{M_\odot}$. APS03 pointed out that an
increase in the angular momentum loss rate at low periods can solve a major
problem regarding the theoretical vs observed values of the period minimum. If
CV evolution at $P < 2$\,h were driven solely by gravitaional radiation,
\citet{pa84} found that the period minimum should be at 1.1\,hr rather than the
observed value of 1.3\,hr. He also noted that the angular momentum loss rates
would be very low for these short-period systems, implying a low mass accretion
rate and therefore a high population of CVs at the minimum period, a prediction
that is contradicted by observation. By adding the extra angular momentum loss
rate from magnetic braking, the cut-off is shifted to longer periods. However,
our value of the magnetic braking rate causes the opposite problem. We find that
at short periods, the magnetic braking rate is almost 100 times the rate of
angular momentum loss due to gravitational radiation. \citet{pa81} showed that, for constant $\dot{J}$, 
\[P_{min} \propto \left(\frac{\dot{J}}{J_{gr}}\right)^{0.34},\]
\citep[see also][]{pa84}, bringing the minimum period up to a value of 331 mins
$\simeq$ 5.5 hours, which is clearly not correct.

We also have the continuing problem of how to explain the presence of the period
gap - a dearth of systems with periods $2\,\rm{h} \leq P \leq 3\,\rm{h}$. If magnetic braking
does not shut off as the secondary becomes fully convective then there is no
reason for systems to cease mass-transfer between those periods. APS03 suggested
that instead of a migration of CVs from long- to short-period, the systems above
and below the period gap may belong to two different populations, with no
migration between the two. However, this is more likely for their
longer-timescale angular momentum loss, as their model depends upon the presence
of an evolved secondary star, than for our measured magnetic braking rate.

\section{Conclusions}

We find that there are two possible explanations for the observed period change
in the pre-CV NN~Ser over the last 15 years. If the change is due to a genuine
angular momentum loss from the system then the rate corresponds to an angular
momentum loss that agrees most closely with the standard magnetic braking rate
proposed by \citet{rvj83}, and that the reduced magnetic braking rate of
\citet{aps03} underestimates the measured rate by $\sim$ 2 orders of magnitude.
We find no evidence for a cut-off in magnetic braking as the secondary mass
drops below $M = 0.3\,\rm{M_{\odot}}$. If the period change is instead due to a third
body, we place constraints on such a body of 0.0043\,M$_{\odot} < M_{3} < 0.18
\,\rm{M}_{\odot}$ and 30 years $<$ P$_{3} <$ 285 years.

As a by-product of this investigation, we have found that the energy
requirements of the \cite{ap92} quadrupolar distortion mechanism are significantly
increased once one accounts for the role of the inner part of the 
star in counter-balancing the outer shell.

\section{Acknowledgements}
CSB was supported by a PPARC studentship. TRM acknowledges the
support of a PPARC Senior Research Fellowship. ULTRACAM is supported
by PPARC grant PPA/G/S/2002/00092. This work is based on observations
from the William Herschel Telescope, operated by the Isaac Newton
Group at the Observatorio del Roque de los Muchachos of the Instituto
de Astrofisica de Canarias. This research was carried out, in part, at the Jet Propulsion Laboratory, California Institute of Technology, and was sponsored by the National Aeronautics and Space Administration.


\bibliography{csb}
\bibliographystyle{../../../Styles/mn2e}

\bsp

\label{lastpage}

\end{document}